\documentstyle[prl,aps,epsfig,floats,twocolumn]{revtex}
\def\ni{\noindent}
\def\br{\vec{r}}
\def\bs{\vec{s}}
\def\bt{\vec{t}}

\def\bR{\vec{R}}

\def\bF{\vec{F}} 
\def\bff{\vec{f}} 
\def\bnabla{\vec{\nabla}}

\def\hC{{\hat C}} 

\def\hP{{\hat P}} 
\def\hG{{\hat G}} 
\def\hQ{{\hat Q}} 
\def\hR{{\hat R}} 
\def\hS{{\hat S}} 
\def\hsigma{{\hat \sigma}} 
\def\hdelta{{\hat \delta}} 
 
\def\cA{{\cal A}} 
\def\cD{{\cal D}} 

\def\halfspace{\hskip0.4cm}

\begin{document}
\draft


\title{The stress field in granular systems: Loop forces and potential formulation}
\author{Robin C. Ball$^1$ and Raphael Blumenfeld$^{1,2}$}

\address{$^1$ Dep. of Physics, University of Warwick, CV4 7AL Coventry, UK.}
\address{$^2$ Polymers and Colloids, Cavendish Laboratory, Madingley Road, 
Cambridge CB3 0HE, UK}
\maketitle
\date{\today}
\maketitle

\begin{abstract}
The transmission of stress through a marginally stable granular  pile in
two dimensions is exactly formulated in terms of a vector field of loop forces, 
and thence in terms of a single scalar potential.

The loop force formulation leads to
a local constitutive equation coupling the stress tensor to {\it fluctuations} in the
local geometry. For a disordered pile of rough grains (even with  simple
orientational order) this means  the stress tensor components are coupled in a
frustrated manner,  analogous to a  spin glass.

If the local geometry of the pile of rough grains has long range staggered 
order,  frustration is avoided and a simple linear theory follows.  
Known exact lattice solutions for rough grains fall into this class.

We show that a pile of smooth grains (lacking friction) can always be mapped into a
pile of unfrustrated rough grains.  Thus it appears that the problems of rough and
smooth grains may be fundamentally distinct.

\end{abstract}
\narrowtext

It has long been recognized in engineering practice that granular materials which
lack cohesion cannot be regarded entirely as solids,  nor (until 'fluidised' or
yielding) can they be regarded entirely as fluids \cite{pw}. 
In this Letter we explore the scenario that there exists a {\it marginal
state} of granular matter in between solid and liquid. 
Most particularly, we are concerned with how that intermediate state transmits stress.

The marginal state is readily characterised in terms of the mean coordination
number $\bar{z}$ for mechanical equilibrium. If external loads,  such as gravity, 
are exerted on a pile of $N$ grains \cite{grain} then the $N\bar{z}/2$ intergranular 
forces must
be able to adjust so as to achieve $Nd(d+1)/2$ constraints balancing the force and
torque on each grain in $d$ dimensions. For perfectly rough grains,  where every
contact supports friction,  this gives a critical coordination number $z_c=d+1$, 
whilst for ideally smooth asymmetric grains with no friction
$z_c=d(d+1)$ \cite{EG},\cite{BGE},\cite{BG}.

For mean coordination numbers
$\bar{z}<z_c$ the pile cannot be stable and under a general loading it must rearrange
or consolidate.  By contrast for $\bar{z}>z_c$ the intergranular
forces are underdetermined by the conditions of force balance alone and the 
deformation of individual grains together with their local constitutive equations
becomes relevant.  In the first case we have granular motion and the pile is 
fluid or yielding,  whereas in the latter case intragranular deformation is
crucial and we have a solid of constitution influenced by that of the individual
grains.  It is interesting to note that for ideally smooth grains, $z_c$ matches the
maximum mean coordination attainable by sequential packing (and in two dimensions
it is the absolute maximum possible \cite{BalltoCates}), so a (random) 
pile of smooth grains is probably never solid.

The marginal case $\bar{z}=z_c$ is special in that the grains do not need to move and 
the intergranular forces,  and hence distribution of stress, are determined by conditions 
of force balance alone.  This has been identified as a paradigm problem of theoretical 
granular mechanics \cite{pw}, \cite{EO}, \cite{paradigm}.  
The macroscopic analogues of balancing force and torque are that the stress tensor 
$\sigma$ obey 

\begin{equation}
\bnabla\cdot\hsigma=\bF \halfspace , \halfspace \cA(\hsigma)=0 
\label{eq:Ai},
\end{equation}
where $\cA(\hsigma)={1\over 2}(\hsigma-\hsigma^T)$ is the antisymmetric part of $\hsigma$ 
and $\bF$ is an external force field. We shall focus here on a region transmitting stress 
but otherwise unloaded.
This system of continuum equations is incomplete without $d(d-1)/2$ further {\it
constitutive relations}.  The central mystery we are seeking to unravel is the
nature of these relations (one relation in two dimensions) when they also arise from
force balance at the granular level.

The key idea of this Letter,  where we limit ourselves for simplicity to the two
dimensional case, is to focus attention on local loops around contact points and
corresponding loop forces. Each local void,  which we will label by an index $l$, 
is enclosed by a loop of grains and contact points;  each contact point separates
two voids and so is party to two loops.  Around each loop we take a {\it
loop force} $\bff_l$ to circulate in the anticlockwise direction,  that is
$+\bff_l$ is contributed to each intergranular force around the loop with
(notionally) positive sign in the anticlockwise direction,  and vice versa for the
reaction forces. The resulting total force across a contact point is then a difference
between two contributing loop forces.

The loop forces have two key features, the first being that they parameterise
the intergranular forces in a manner that automatically satisfies balance of force 
at each contact point and on each grain.  
In terms of them we can write the stress tensor on each grain as

\begin{equation}
\hS_g = \sum_l{\br_{lg} \bff_l }
\label{eq:Aii}
\end{equation}
where $\vec{r}_{lg}$ is the vector connecting (anticlockwise) the two contact points
shared by loop $l$ and grain $g$ and $A_g$ is the grain area defined in fig. 1 
\cite{Ag}.  
The vectors $\sum_g{\vec{r}_{lg}} = 0$ form an
anticlockwise loop round the contact points of loop $l$ and likewise 
$\sum_l{\vec{r}_{lg}} = 0$ forms a clockwise loop round the contact points of grain $g$
and play a central geometrical role in our discussion.  

The second key feature of the loop forces is that, being defined on the loops 
which are half as numerous as the grains, they are a comparatively coarse-grained 
quantity. The hidden nature of the constitutive equation can now readily be
appreciated,  because the only remaining constraints that the loop forces have to obey
is balance of torque around each grain, that is

\begin{equation}
\sum_l \br_{lg} \times \bff_l = 0
\label{eq:Aiii}
\end{equation}
which is one equation per grain and hence {\it two} equations per loop.

In essence the two torque equations per loop give us two macroscopic conditions, 
$\cA(\hsigma)=0$ {\it plus} a constitutive relation.  To achieve this separation
explicitly we postulate a smooth interpolation of the loop forces to a function 
of continuous position
$\bff(\br)$,  with each loop and grain having a nominal centre $\bR_l$ and 
$\bR_g$, respectively. This also introduces a natural continuum already at the 
grain scale without going into elaborate coarse-graining schemes.
Then the force moment on grain $g$ is, to a first order Taylor
approximation,

\begin{equation}
\hS_g = \hC_g \cdot \bnabla \bff
\label{eq:Aiv}
\end{equation}
where

\begin{equation}
\hC_g = \sum_l \br_{lg} \bR_{lg} \halfspace \hbox{and} \halfspace
\bR_{lg}=\bR_l-\bR_g
\label{eq:Av}.
\end{equation}

The geometric tensors $\hC_g$ characterise the grain local geometry and their symmetric and antisymmetric parts have quite distinctive properties. It is readily shown that 

\begin{equation}
\hC_g = A_g\hR +  \hP_g 
\label{eq:Avi}
\end{equation}
where $\matrix{\hR}={0 \,1 \choose -1 \,0 }$ is the unit antisymetric tensor (also
corresponding to ${\pi\over 2}$-rotation in the plane) and $A_g$ is the area associated with grain $g$ shaded in figure 1.  The areas $A_g$ have the convenient property of tessalating the plane. 
The symmetric part, $\hP_g$, reflects {\it fluctuation in} the departure of the local geometry  from
isotropy,  being given by
\begin{equation}
\hP_g = \sum_l{ \frac{1}{2}(\bs_{lg} \bs_{lg} - \bt_{lg} \bt_{lg})  }
\label{eq:Avi.1}
\end{equation}
where the vectors $\bs_{lg}$ and $\bt_{lg}$ are shown on figure 1.  
It is a crucial point that neighbouring grains share equal and opposite
contributions $\bs \bs$ and $-\bt \bt$ to their $\hP$ tensors,  
and from this it follows that the average $\langle \hP \rangle_G$ of $\hP$ over
 all the grains in any convex region $G$ goes to zero at least as fast as $N_G^{-1/2}$,  
where $N_G$ is the number of grains in $G$.  Note that the averaging of the $P_g$ does not depend on any assumptions of isotropy.
For periodic lattices  $\langle \hP \rangle$ vanishes over one unit cell.

To underpin the significance of $C_g$ as the relevant geometrical order parameter, consider a region of area $A$ 
whose boundary is defined by the contact points of the outermost grains. Averaging over $\hS_g$ in 
the region using eq. (\ref{eq:Aiv}) gives 

\begin{equation}
\hsigma = {1\over A}\sum_g \hC_g\cdot\bnabla\bff = 
\langle \hC_g\cdot\bnabla\bff\rangle \ .
\label{eq:Bii}
\end{equation}
Carrying out the same summation using eq. (\ref{eq:Aii}) we observe 
that the sum over loops inside the region cancels out and the only 
contribution comes from the boundary vectors, $\br_b$, between 
boundary contact points, 

\begin{equation}
\hsigma = 
{1\over A}\sum_{\rm boundary} \br_b\bff_b \ ,
\label{eq:Biii}
\end{equation}
where $\bff_b$ is the external loading on the region. This contour sum can be converted, 
using Stokes theorem, into 

\begin{equation}
\hsigma = {1\over A}\int ds \hR\cdot\bnabla\bff = 
\hR\cdot\langle\bnabla\bff\rangle \ .
\label{eq:Biv}
\end{equation}
Since $\langle\hP_g\rangle=0$ we identify $\langle\hC_g\rangle = A \hR$ 
and, on comparing to eq. (\ref{eq:Bii}), we obtain

\begin{equation}
\langle\hC_g\cdot\bnabla\bff\rangle =  
\langle\hC_g\rangle\cdot\langle\bnabla\bff\rangle \ .
\label{eq:Bi}
\end{equation}
This result is particularly reassuring because it yields the effective response 
of a macroscopic granular region to a force field in terms of an effective 
macroscopic characteristic property $\langle\hC_g\rangle$. It is equivalent 
to results in contexts such as disordered dielectrics and 
continuum elasticity that are derived by combining 
a field equation with either a constitutive relation or an energy functional.

We can now construct the constitutive equation in terms of the continuum stress tensor,  which is given from eq. (\ref{eq:Biv}) as $\hsigma = \cD \bff$ where 
$\cD$ is exactly the curl operator in two dimensions.  
The condition that the stress tensor on each grain be symmetric then leads,  
over and above the condition that the continuous $\hsigma$ be symmetric, to the 
requirement

\begin{equation}
{\rm Tr}(\hQ\cdot\hsigma)=0
\label{eq:Aviii},
\end{equation}
where $\hQ=\hR\cdot\hP\cdot\hR^T$ is a rotated verion of the symmetric part of the 
local $\hC_g$ tensor. 
Relation (\ref{eq:Aviii}) is the continuum constitutive equation that provides the 
missing link between the stress and the local geometry. 
It has the striking feature that the coefficients $\hQ$ are
spatially fluctuating quantities that locally add to zero, so that any simple attempt
to identify a non-vanishing mean field value $\langle\hQ\rangle$ leads back to Eq. 
(\ref{eq:Ai}) and therefore yields no new constitutive information.
In this sense the problem is analogous to spin glasses, in our case it being 
$\hsigma$ that is subject to spatial couplings of random sign \cite{commentEG}.

Constitutive equations coupling stress components in a local linear relation,  
represented here by (\ref{eq:Aviii}), have been discussed before under the 
presumption that a non-vanishing macroscopic version of $\hQ$ exists.  
Examples are the hypotheses of "fixed principal axes" \cite{catesi} 
(corresponding to $\hQ$ being traceless) and 
"oriented stress linearity" \cite{catesii} that were introduced to describe 
sand piles \cite{EO}. Having shown that the simple grain average of $\hQ$ 
vanishes, we are therefore particularly motivated to consider any case where a 
non-zero effective $\hQ$ can be identified.

Drawing inspiration from spin problems,  we focus on the analogy of an 
antiferromagnet in which the local $\hP$ tensors have alternation of sign in an 
unfrustrated pattern.  The key idea is that for systems where every loop has an 
even 
number of edges, we can label all grains  
+ or - such that each grain is surrounded only by 
opposite sign neighbours in an unfrustrated way. Then we can distinguish locally 
averaged values $\hP_+$ and $\hP_-$ restricted to the respective $+$ and $-$ 
grains,  where $\hP_+ + \hP_- = 0$.  If the pile has staggered geometrical order 
such that these are non-zero,  then this leads to the identification 
\begin{equation}
\hQ_{{\rm eff}}=\hR \hP_+ \hR^T
\label{eq:Do}
\end{equation}
The simplest such case where we can see this work is a periodic lattice (fig. 2) 
with a generally anisotropic unit 
cell comprising of two grains,  which we duly label $+$ and $-$. 
Using eq. (\ref{eq:Aiv}) for each grain separately, we write down the mean and 
deviatory cell stresses, for which we find

\begin{equation}
\hsigma = \left( \hS_+ + \hS_- \right) = \left( A\hR + \hP_++\hP_- 
\right)\cdot\bnabla\bff 
\label{eq:Di}
\end{equation}
and
\begin{equation}
\hdelta = \left( \hS_+ - \hS_- \right) = 
\left( \delta A\hR + \hP_+-\hP_- \right)\cdot\bnabla\bff \ ,
\label{eq:Dia}
\end{equation}
where $A$ is the unit cell area and $\delta A = A_+-A_-$ is the difference 
between the 
areas associated with the two grains. 
Using the periodic properties of the lattice one readily obtains
\begin{equation}
\hP_+ = -\hP_- = \frac{1}{2}\sum_k{\bs_{k} \bs_{k}-\bt_{k} \bt_{k}}
\label{eq:Dib}
\end{equation}
where the vectors $\bs_{k}$ and $\bt_{k}$ are labelled on figure 2.
Then the condition that both $\hdelta$ as well as $\hsigma$ be symmetric then leads to 
the constitutive equation (\ref{eq:Aviii}) with $\hQ$ replaced according to (\ref{eq:Do}).

The idea of a staggered order parameter can be further exploited 
to formulate a mapping from perfectly smooth to perfectly rough systems. 
Given a pile of perfectly smooth grains, we note that its behaviour is 
identical to the same pile of rough grains provided at each contact point 
we insert a vanishingly small perfectly rough ball bearing. It is readily 
checked that starting from $\bar{z}=6$ for the smooth grains leads to 
$\bar{z}=3$ for the resulting set of rough grains (3/4 of which are the new 
bearings each having $z=2$),  so we duly arrive at a marginally connected set of 
rough grains.  Furthermore, we automatically have an alternating distinction 
between original grains ($+$) and bearings ($-$).  Explicit evaluation shows 
that for each small grain,  $\hP_- \propto \epsilon$ as bearing radius $\epsilon 
\rightarrow 0$, so in this case the appropriate geometrical order parameter is 
$p=Limit_{\epsilon\rightarrow 0} P_+/\epsilon$.  This order parameter,  and hence the 
corresponding $Q_{{\rm eff}}$, can have a non-zero average for an 
orientationally ordered assembly of smooth grains, as computed explicitly for a 
periodic array\cite{BG}.  This contrast to the case of 
general rough grain piles,  which explicitly require staggered orientational order 
for a non-zero $Q_{{\rm eff}}$,  suggests that piles of perfectly smooth grains may be 
inherently distinct from piles of rough ones.

Finally we introduce an alternative formulation of the problem where the loop forces
are defined in terms of a potential field. Since each loop is associated  with a
two-dimensional force the number of degrees of freedom that the  potentials should
provide is twice the number of loops, which (up to relative  correction of
$1/\sqrt{N}$ due to the boundary) is exactly the number of grains in the system.
Thus, we require that the potential field provide one degree of freedom per grain
leading to scalar grain potentials, $\psi_g$.  A natural way to precisely define the
potentials is to choose that the forces are derived from them via

\begin{equation}
\bff_l = {{-1}\over{A_l}} \sum_g \br_{lg} \psi_g \ ,
\label{eq:Ci}
\end{equation}
and the potentials are then governed by eq. (\ref{eq:Aiii}).

We now assume that the $\psi$-field can be continued throughout the system
just as we did previously for the force field.  Then expanding around the loop
centres, we obtain

\begin{equation}
\bff_l = {1\over{A_l}} \hC_l\cdot \bnabla\psi \ .
\label{eq:Cii}
\end{equation}
The geometrical tensor $\hC_l = \sum_g \br_{lg} \bR_{lg}$ is the loop 
analogue of $\hC_g$ and its antisymmetric part is likewise associated with an
 area $A_l$,  where these areas tessalate the plane.  The symmetric part of 
$\hC_l$ is sensitive to the choice of grain centre positions:  as there are
two grains per loop these constitute four degrees of freedom per $\hC_l$ 
tensor and the expectation is that these tensors can thereby be chosen to be 
purely antisymmetric.  Using $\hC_l = A_l \hR$ simplifies the expression of
the local stress in terms of the potentials to

\begin{equation}
\hS_g = \hC_g\hR^T\cD_g\cD_l\psi 
\label{eq:Ciii}
\end{equation}
and the continuum stress becomes

\begin{equation}
\hsigma = \cD_g\cD_l\psi \to {\rm {\bf curl\ curl}}\psi \ .
\label{eq:Civ}
\end{equation}
It follows that the continuous version of $\psi$ is exactly the Airy stress function 
\cite{Muskh}.

To conclude, we have formulated a theory of stress transmission in granular matter 
at the marginal state between solid and fluid in terms of loop forces, which in 
turn can be defined in terms of a scalar potential. 
Granular systems can possess a predominant antiferromagnetic order but with local frustration, 
much like in antiferromagnetic Ising systems. An antiferromagnetic order 
yields a global consitutive relation leading to a macroscopic theory while frustration 
requires a more sophisticated approach.
The similarity between random granular piles and spin glasses suggests a 
possible use of methods from the latter to this problem. We have mapped smooth piles to 
antiferromagnetically ordered perfectly rough ones obtaining for them an exact constitutive 
relation. This result points to an inherent 
difference between rough (frustrated) and smooth (ordered) systems and may thus undermine the 
idea of gaining insight into behaviour of piles of finite friction grains from understanding 
the perfectly smooth and perfectly rough limits. 

An extension of this work to three dimensions is currently under way.

\ni {\bf Acknowledgements}

\ni We are grateful to Prof Sir S. F. Edwards and Dr D. Grinev for many 
illuminating discussions.  This research was supported by EPSRC grant 
GR/L59757.

\begin{figure}
\centerline{\psfig{file=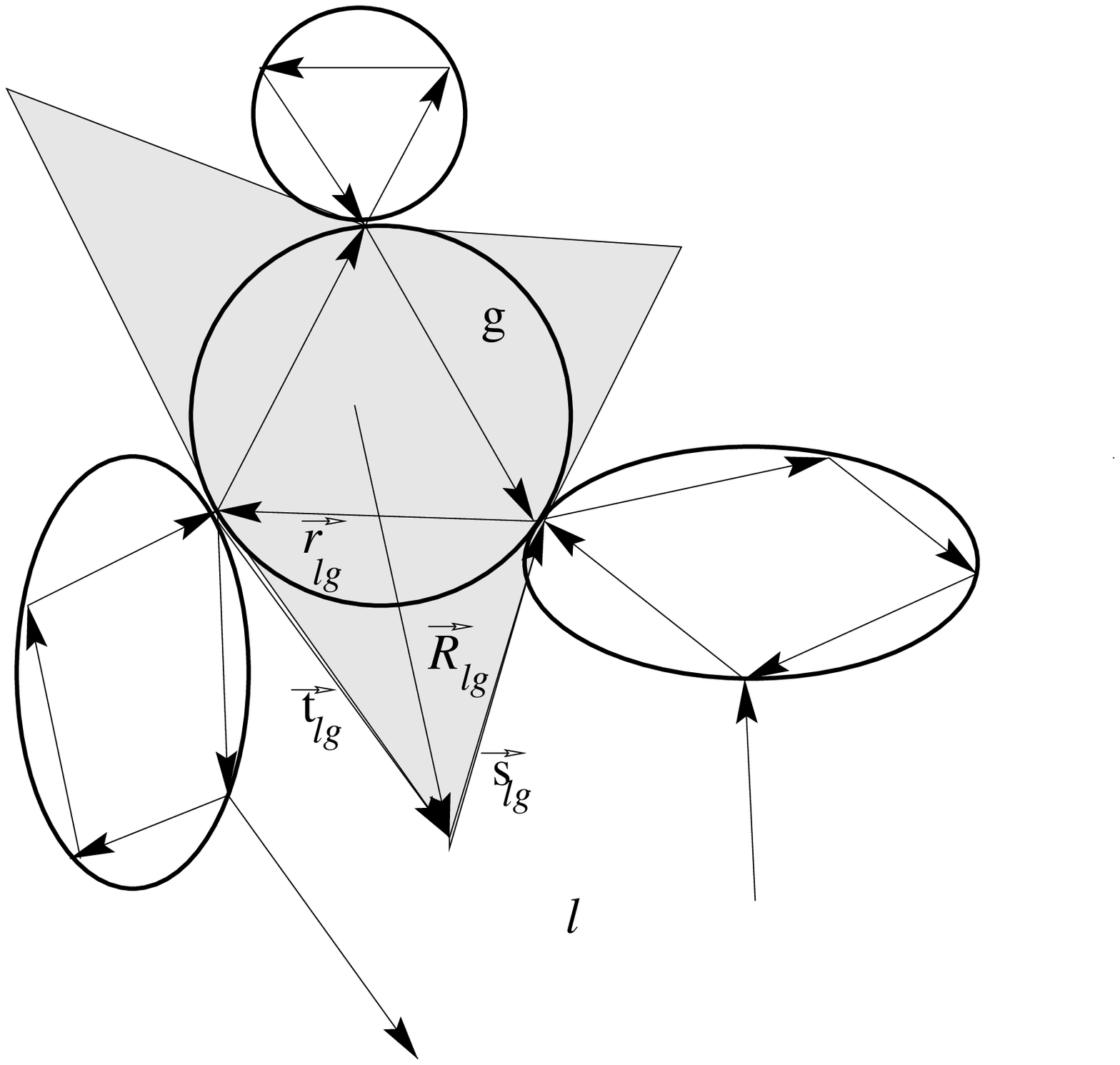,height=8.0cm}}
\caption{
The local geometry around a grain $g$.  The vectors $\br_{lg}$ connect contact 
points clockwise around each grain $g$ and anticlockwise around each void loop 
$l$,  whilst the vectors $\bR_{lg}$ connect from grain centres to loop centres.  
The stress field is coupled through tensors $\hC_g = \sum_l{\bR_{lg} \br_{lg}}$,  
whose antisymmetric part has magnitude $A_g$ which is the area shaded in the 
figure.  The symmetric parts of these tensors evaluate to 
$\hP_g=\frac{1}{2}\sum_l{\bs_{lg} \bs_{lg}-\bt_{lg} \bt_{lg}}$ and play a crucial 
role in the constitutive equation.  As can be seen from the labelling of the 
vectors $\bs_{lg}$ and $\bt{lg}$ in the figure,  neighbouring grains share equal 
and opposite contributions to their repsective $\hP_g$,  so these tensors tend 
to zero upon local averaging.}
\end{figure}
\begin{figure}
\centerline{\psfig{file=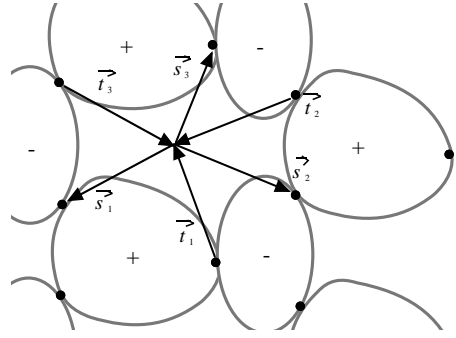,height=6.5cm}}
\caption{Part of an anisotropic periodic lattice,  where there are two distinct grains, 
labelled + and -,  per unit cell.  Stress transmission is controlled by the tensor 
$\hP_+ = -\hP_- = \frac{1}{2}\sum_k{\bs_{k} \bs_{k}-\bt_{k} \bt_{k}}$.  The
lattice vectors are $\bt_1+\bs_3$, $\bt_2+\bs_1$ and $\bt_3+\bs_2$ and because these are
not independent the value of $\hP_+$ turns out to be independent of the choice of loop centre.}
\end{figure}


\begin{thebibliography}{99}
\bibitem{grain} A grain is defined as a rigid
particle of finite volume and arbitrary shape.
\bibitem{Ag} The definition of $A_g$ as a grain area
is useful because the entire system can be uniquely mapped in
terms of these areas and $A_{system}=\sum_g A_g$.
\bibitem{commentEG} The conclusion that the stress couplings have zero average appears 
to be at variance 
with that of \cite{EG} who arrived at an equation similar (but not identical)  
to equation (\ref{eq:Aviii}).  However, we find that their coupling tensors $\hG^{\alpha}$ 
have similar averaging properties to our $\hP_g$.
\end{thebibliography}
\end{document}